\documentclass[reqno]{amsart}

\usepackage{amsmath}
\usepackage{amsfonts}
\usepackage{amsopn}
\usepackage{amsthm}
\usepackage{graphics}
\usepackage[dvips]{graphicx}
\usepackage{subfig}

\begin{document}

\title[Enhanced FDTD Method for Maxwell's Equations]
{An Enhanced Finite Difference Time Domain Method for Two
Dimensional Maxwell's Equations}

%\subjclass{Primary 65M06, 65M12}

%\date{October 31, 2019}

\author{Timothy Meagher$^{1}$, Bin Jiang$^{2}$, Peng Jiang$^{3}$}

\footnotetext[{1}]{Department of Mathematical Sciences, University
of Nevada, Las Vegas, NV 89154, USA}

\footnotetext[{2}]{Fariborz Maseeh Department of Mathematics and
Statistics, Portland State University, Portland OR 97201, USA}

\footnotetext[{3}]{Department of Chemical Engineering, University of
Florida, Gainesville, FL 32611, USA}

\bibliographystyle{unsrt}

\begin{abstract} An enhanced finite-difference time-domain (FDTD) algorithm
is built to solve the transverse electric 2D Maxwell's equations
with inhomogeneous dielectric media where the electric fields are
discontinuous across the dielectric interface. The new algorithm is
derived based upon the integral version of the Maxwell's equations
as well as the relationship between the electric fields across the
interface. To resolve the instability issue of Yee's scheme
(staircasing) caused by discontinuous permittivity across the
interface, our algorithm revises the permittivities and makes some
corrections to the scheme for the cells around the interface. It is
also an improvement over the contour-path effective permittivity
algorithm by including some extra terms in the formulas. The scheme
is validated in solving the scattering of a dielectric cylinder with
exact solution from Mie theory and is then compared with the above
contour-path method, the usual staircasing and the volume-average
method. The numerical results demonstrate that the new algorithm has
achieved significant improvement in accuracy over other methods.
Furthermore, the algorithm has a simple structure and can be merged
into current FDTD software packages easily. The C\texttt{++} source
code for this paper is provided as supporting information for public
access.
\end{abstract}


\maketitle

\section{Introduction}

The finite-difference time-domain (FDTD) algorithm is one of the
most popular numerical methods to solve Maxwell's equations, as
proposed by Yee \cite{Yee1966}. It is a time marching forward method
which easily makes a visual representation of the fields. It also
scales very well since the number of computation required is
proportional to the size of the model and the method requires no
large-scale linear algebra computation \cite{Taflove2005}. The FDTD
method requires a completely structured grid also known as the Yee
grid. However, Many real world problems have complicated geometries
and numerical inaccuracy happens when the modeled objects do not fit
the grid very well.

In this paper, we focus on two-dimensional transverse electric (TE)
Maxwell's equation within multiple non-magnetic media where the
electric fields are not continuous across the media interface so the
numerical schemes need to be designed to take care of such
discontinuity. The other case is the transverse magnetic equation
where the electric field is always continuous across the media
interface and the original algorithm will work as in the homogeneous
medium.

The original FDTD algorithm proposed in Yee's paper assigns each of
the field component an electric permittivity and a magnetic
permeability solely based on the material properties at its
location. This is commonly known as the staircasing or pixelation
method. In general, the staircasing method has an error that scales
with $\mathcal{O}(\Delta x^2)$ given a grid size of $\Delta x$ for
cells that are homogenous. However, in cells that contain a medium
interface, the local error becomes $\mathcal{O}(1)$. Even though the
number of cells that contain an interface is often a small fraction
of the total cells, these local errors can cause a rough global
error of $\mathcal{O}(\Delta x)$.

Subsequent research in modern optics and electromagnetism dealing
with the Maxwell's equations have focused on more and more complex
geometries with different materials so that the complicated
dielectric tensor can cause serious pixelation issues. In turn,
there have been many improvements in FDTD algorithms to overcome
these pixelation issues. One way to overcome the pixelation problem
is by fitting the mesh to the device, thereby generating the
so-called non-orthogonal FDTD \cite{Harms1992, Liu2009}. The second
method is to refine the Cartesian mesh around the interfaces and is
thus called subgridding method \cite{Zivanovic1991}. Even though
these two methods converge faster than the standard FDTD method,
their complexity are larger than original FDTD method and they also
have time stability issues.

In order to keep a completely structured grid and maintain all the
benefits of FDTD, another method redefines Maxwell's integral
equations around the interface of the media. This is known as the
effective-permittivity (EP) method. This method uses sub-pixel
smoothing techniques to change the permittivities of the field
components around the interface to produce better results by taking
account of many different factors, such as the interface conditions,
permittivities of adjacent field components and so on. The revised
effective permittivities smooth out the pixel error. The main
benefit of these methods is that there is a negligible increase of
the numerical load since all the revised permittivities have been
calculated before the main FDTD algorithm loop.

The first method of sub-pixel smoothing is the volume average
effective permittivity (V-EP) \cite{Dey1999}. This method assigns
effective permittivities for all the field components in a cell that
contains media interface. The effective permittivities are
calculated by taking a weighted average based on how much percent of
the volume each medium occupies. This method is simple for
implementation and also stabilizes the error fluctuation that occurs
in the staircasing method. However, it does not decrease the error
rapidly and its error can be worse than staircasing method sometimes
due to ignorance of the interface orientation.

The second method proposes dynamic formulas for the EP based on the
orientation of the media interface and their expressions are
accurate when the interface is perpendicular or parallel to the mesh
axes \cite{Kaneda1997, Hwang2001, Liuchang2012}. These EPs improve
the accuracy of the FDTD method, while keeping the simple structure
of the original algorithm. However, their performance will
deteriorate for a curved or flat interface not perpendicular or
parallel to the mesh axes.

The third method uses the reflection coefficients to derive
effective permittivities and shows the method can also achieve
second-order convergence with several special slanted angles between
the interface and the Yee grid rather than just orthogonal or
parallel to the Yee grid \cite{Hirono2000, Hirono2010}. However, its
derivation is quite complicated and has no ways to be extended to
arbitrary interfaces.

In a landmark paper \cite{Mohammadi2005}, Mohammadi et al. proposed
the so-called contour-path effective permittivities (CP-EP) method.
The CP-EP method first addresses the orthogonal and parallel cases
as shown above, then extends the idea to handle with more general
geometry. Unlike the previous methods, it could handle arbitrary
boundary with any orientation of the boundary. Compared with
traditional staircasing methods, CP-EP has almost no additional
runtime cost since determination of the effective permittivities has
been done before the main FDTD loop.

However, we will demonstrate that CP-EP does not incorporate one
important term from the interface conditions into the FDTD
algorithm. Then we will develop a revised EP algorithm where the
missing term will be re-considered in order to produce a more
accurate and stable algorithm. Since this method incorporates the
boundary conditions very well, it will be denoted as BC-EP.

In fact, such deficiency of CP-EP has also been identified in
\cite{Farjadpour2006, Oskooi2009} where effective dielectric
permittivity tensors are constructed to take care of both isotropic
and anisotropic materials so as to achieve second-order convergence
under the open source software MEEP. However, such proposed method
which satisfies the interface conditions for electromagnetic fields
has been shown to have late-time instabilities, and many
possibilities to average the effective dielectric tensor are
explored to avoid late time instabilities \cite{Werner2007}. Further
efforts have been reported in \cite{Bauer2011} to construct a new
second-order scheme by taking the average of eight triplets.
Although the new method is highly accurate, its effective dielectric
permittivity tensor can still be asymmetric thus unstable for
certain conditions. Therefore, the last improvement of the effective
dielectric tensor has been given in \cite{Werner2013} to make it
symmetric and stable. The numerical results show that this scheme
gives the best result in general and the error in practice still
lies in between first and second-order in most cases.

In the following, we will introduce the new BC-EP method by adding a
few terms to the established CP-EP method for the cells around the
interface so as to keep the numerical load increase as small as
possible and improve the accuracy significantly. Then the numerical
tests verify that BC-EP has a much better performance than CP-EP,
V-EP and staircasing methods while still maintaining numerical
stability. BC-EP has a very simple structure and can be merged into
any existing FDTD packages easily, compared with other established
FDTD software \cite{Oskooi2009, Werner2013} which can be run only
within their framework. We provide C\texttt{++} source code for
BC-EP together other three methods as supporting information for
public access. Conclusion and some future research consideration
will be provided in the end.

\section{New Algorithm Design}

In this section, we will propose a new effective permittivity method
to solve Maxwell's equations in a domain composed of two different
non-magnetic media with different electric permittivities. To simply
the algorithm description, we will focus on the most important
transverse electric (TE) scenario. That is,
$\mathbf{E}(x,y)=E_x\mathbf{i}+E_y\mathbf{j}$ is located in the
incident xy-plane while $\mathbf{H}(x,y)=H_z\mathbf{k}$ is along
z-axis. Moreover, we restrict the discussion to dielectric and
non-magnetic media such that $\mathbf{D}(x,y)=\epsilon
\mathbf{E}(x,y)$ and $\mathbf{B}(x,y)= \mathbf{H} (x,y)$ where
$\epsilon$ stays as a constant in different media and
$\epsilon_0=\mu_0=1$. Based on the integral version of the Maxwell's
equations, we will derive the new algorithm for any orientation of
the interface. This is a significant improvement over all current
methods which are only accurate under specific angles between the
interface and Cartesian coordinate system, such as parallel or
orthogonal to Yee-axis.

\begin{figure}[h]
    \begin{center}
            \includegraphics[scale=0.2]{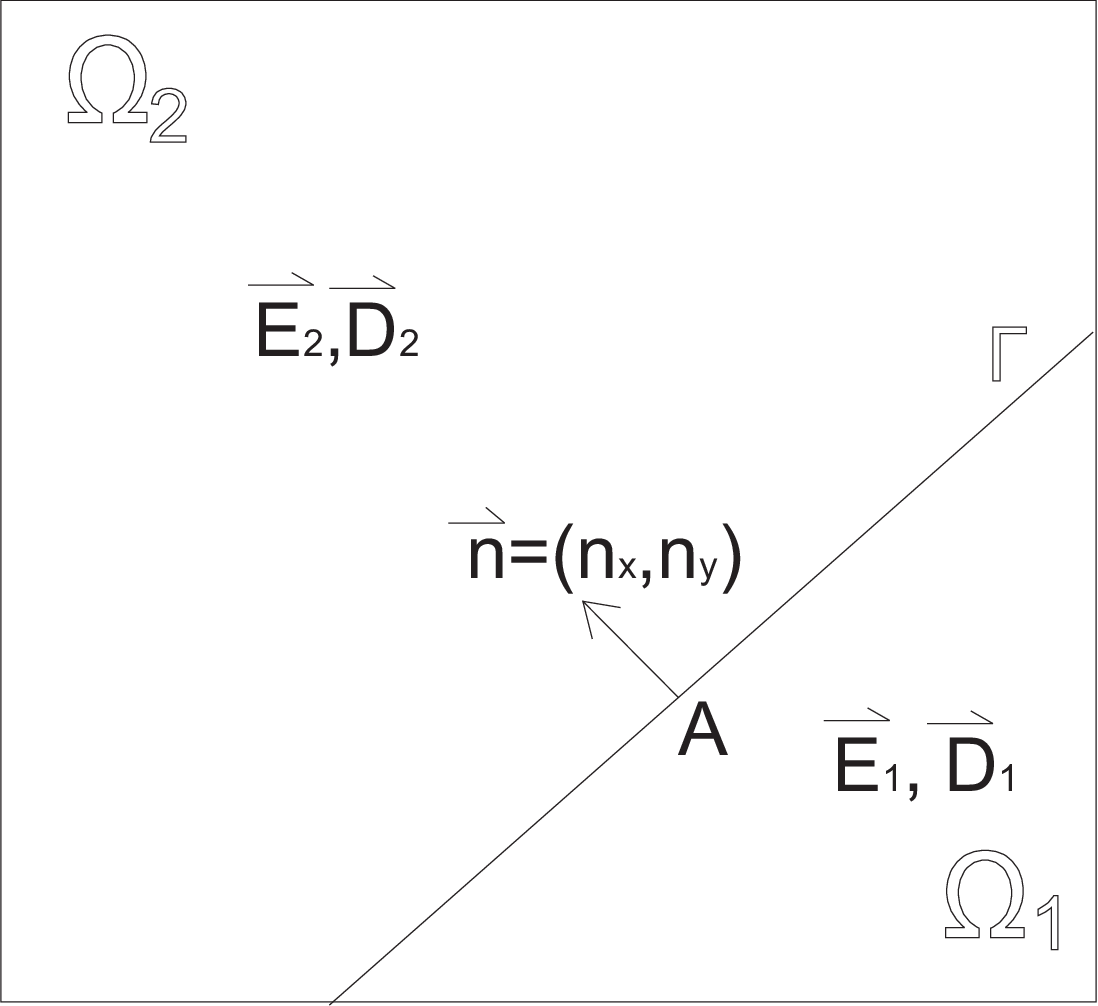}
                \caption{Relation of $\mathbf{D}$ and $\mathbf{E}$ across interface $\Gamma$}
                \label{interface}
    \end{center}
\end{figure}

As a preliminary step, let us set up the relation of
$\mathbf{D}(x,y)$ and $\mathbf{E}(x,y)$ across the interface of
these two media. It is well known that $\mathbf{H}(x,y)$ is
continuous in the whole domain. Meanwhile, across the media
interface, the tangential component $\mathbf{E}_\tau(x,y)$ of the
electric field and the normal component $\mathbf{D}_n (x,y)$ of the
electric flux are continuous as well, as shown in
Figure~\ref{interface} where $\mathbf{n}=(n_x, n_y)$ is the outward
unit norm vector from region 1.
Choose $A \in \Gamma$. $\mathbf{D}_1 (A)=(D_{x1}(A), D_{y1}(A))$ and
$\mathbf{D}_2 (A)=(D_{x2}(A), D_{y2}(A))$ are the electric fluxes at
A from different media, and $\mathbf{E}_1(A)=(E_{x1}(A), E_{y1}(A))$
and $\mathbf{E}_2(A)=(E_{x2}(A), E_{y2}(A))$ are the corresponding
electric fields.

Suppose $\mathbf{n}=(n_x, n_y)$ is the unit norm vector at $A$ from
region 1, then the corresponding tangent unit vector is
$\mathbf{\tau}=(n_y, -n_x).$ Therefore, we can calculate $E_\tau$
and $D_n$ at point $A$ in different regions as follows.
\begin{align}
   E_{\tau_1}= \mathbf{E}_1 \cdot  \mathbf{\tau}=E_{x_1}n_y-E_{y_1}n_x
\end{align}
\begin{align}
   E_{\tau_2}= \mathbf{E}_2 \cdot  \mathbf{\tau}=E_{x_2}n_y-E_{y_2}n_x
\end{align}
Meanwhile,
\begin{align}
   D_{n_1}= \epsilon_1 \mathbf{E}_{n_1}=\epsilon_1 E_{x_1}n_x+ \epsilon_1E_{y_1}n_y
   \label{Dinterface}
\end{align}
\begin{align}
   D_{n_2}= \epsilon_2 \mathbf{E}_{n_2}=\epsilon_2 E_{x_2}n_x+ \epsilon_2 E_{y_2}n_y
\end{align}
Based on the fact that $E_{\tau_1}=E_{\tau_2}$ and
$D_{n_1}=D_{n_2}$, we have
\begin{align*}
    \begin{cases}
    E_{x_1}n_y-E_{y_1}n_x & =E_{x_2}n_y-E_{y_2}n_x\\
    \epsilon_1  E_{x_1}n_x+\epsilon_1E_{y_1}n_y &= \epsilon_2 E_{x_2}n_x+\epsilon_2 E_{y_2}n_y
    \end{cases}
\end{align*}
Therefore, by using $n_x^2+n_y^2=1$, we get
\begin{align*}
    \begin{bmatrix}
        E_{x_1}   \\
    E_{y_1}
     \end{bmatrix}=
      \begin{bmatrix}
        n_y &  - n_x    \\
        \epsilon_1 n_x   &  \epsilon_1  n_y
     \end{bmatrix} ^{-1}
    \begin{bmatrix}
        n_y &  - n_x    \\
        \epsilon_2 n_x   &  \epsilon_2  n_y
     \end{bmatrix}
     \begin{bmatrix}
        E_{x_2}   \\
    E_{y_2}
     \end{bmatrix}
\end{align*}
Or
\begin{align}
    \begin{cases}
    E_{x_1} & = (\frac{ \epsilon_2}{ \epsilon_1}n_x^2+n_y^2) E_{x_2} + (\frac{ \epsilon_2}{ \epsilon_1} -1)n_xn_yE_{y_2}\\
    E_{y_1} & = (\frac{ \epsilon_2}{ \epsilon_1} -1)n_xn_y E_{x_2}   + (n_x^2+\frac{\epsilon_2}{\epsilon_1}n_y^2)E_{y_2}
     \end{cases}
     \label{eintro}
\end{align}
Furthermore, by using $\mathbf{D}_i=\epsilon_i\mathbf{E}_i$ at $A$,
we obtain
 \begin{align}
    \begin{cases}
    D_{x_1} & = (\frac{\epsilon_1}{\epsilon_2}n_y^2+n_x^2) D_{x_2} + (1-\frac{\epsilon_1}{\epsilon_2})n_xn_y D_{y_2}\\
    D_{y_1} & = (1-\frac{\epsilon_1}{\epsilon_2})n_xn_y D_{x_2} + (\frac{\epsilon_1}{\epsilon_2}n_x^2+n_y^2) D_{y_2}
     \end{cases}
     \label{dintro}
\end{align}
If the interface $\Gamma$ is parallel or orthogonal to $x$-axis,
either $n_y=0$ or $n_x=0$ so $n_x \cdot n_y=0$. Then the relation
\eqref{eintro} involving $\mathbf{E}_1$ and $\mathbf{E}_2$ will be
simplified such that $E_{x_1}$ and $E_{y_1}$ will only depend on
$E_{x_2}$ or $E_{y_2}$, respectively. Same holds for $\mathbf{D}_1$
and $\mathbf{D}_2$ in relation \eqref{dintro}.

By using \eqref{dintro}, we can express $D_{x_2}$ in region 2 based
on its neighbor $D_{x_1}$ in region 1 and its corresponding
$D_{y_2}$ in the same region 2. Similar formula can also be derived
for $D_{y_2}$ based on $D_{y_1}$ and $D_{x_2}$.
\begin{align}
    D_{x_2} & = \frac{\epsilon_2}{\epsilon_2 n_x^2+\epsilon_1 n_y^2}D_{x_1}
                 + \frac{(\epsilon_1-\epsilon_2)n_xn_y}{\epsilon_2 n_x^2+\epsilon_1 n_y^2}D_{y_2} \label{dintroDx} \\
    D_{y_2} & = \frac{(\epsilon_1-\epsilon_2)n_xn_y}{\epsilon_1 n_x^2+\epsilon_2 n_y^2}D_{x_2}
                 + \frac{\epsilon_2}{\epsilon_1 n_x^2+\epsilon_2 n_y^2}D_{y_1} \label{dintroDy}
\end{align}
\eqref{eintro} will be used to derive the new scheme for Faraday's
law while \eqref{dintroDx} and \eqref{dintroDy} will be used to
derive the new scheme for Ampere's law.

Now let us express Ampere's law in integral form
\begin{align}
   \frac{\partial }{\partial t} \iint_S \mathbf{D} \cdot  \mathbf{n} dS= \oint_{\partial S}  \mathbf{H} \cdot d\mathbf{l}
   \label{Ampere's}
\end{align}
and Faraday's law integral form
\begin{align}
   \frac{\partial }{\partial t} \iint_S \mathbf{B} \cdot \mathbf{n} dS=- \oint_{\partial S}  \mathbf{E} \cdot d\mathbf{l}
   \label{Faraday's}
\end{align}

Firstly, let us discretize Ampere's law \eqref{Ampere's} around the
interface $\Gamma$ to update $E^n_x$ and $E^n_y$ as below.
\begin{figure}[ht!]
    \begin{center}
            \includegraphics[scale=0.2]{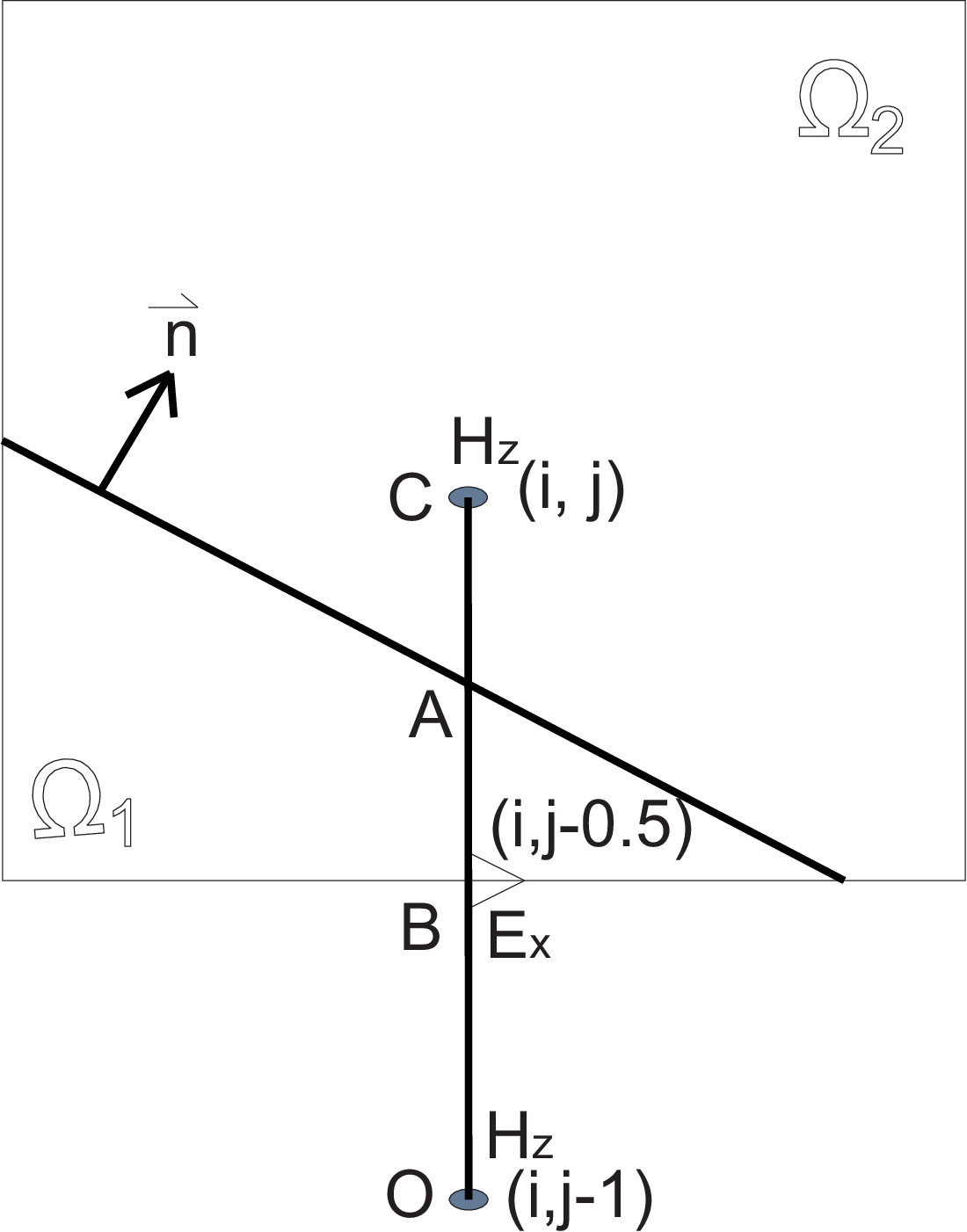}
                \caption{Discretization of Ampere's law across $\Gamma$}
                \label{amex}
    \end{center}
\end{figure}
To update $E_x$, we choose $S$ in \eqref{Ampere's} to be the
rectangle in $yz$-plane whose projection in $xy$-plane is the line
segment $OC$. Suppose the interface $\Gamma$ intersect $OC$ at point
$A$ and $OA=f$ as in Figure~\ref{amex}. Applying time discretization
to Ampere's law along $S$ at time $t_{n+\frac{1}{2}}$ yields
\begin{align}
    \frac{\int_O^C D^{n+1}_{x_{(i,y)}}dy-\int_O^C D^{n}_{x_{(i,y)}}dy}{\Delta t}=H^{n+\frac{1}{2}}_{z_{(i,j)}}-H^{n+\frac{1}{2}}_{z_{(i,j-1)}}
    \label{newdisAmpere's}
\end{align}
Based on formula \eqref{dintroDx}, we can compute the integrals of
left hand side as follows. From now on, all the higher-order terms
of $dx$ and $dy$ are ignored for the clarification of the formulas.
\begin{align}
    \int_O^C D^{n+1}_{x_{(i,y)}}dy &= \int_O^A D^{n+1}_{x_{(i,y)}}dy+\int_A^C D^{n+1}_{x_{(i,y)}}dy \nonumber \\
    &= f D^{n+1}_{x_{(i,j-\frac{1}{2})}} + (\Delta y - f) D^{n+1}_{x_{(A,\Omega_2)}} \nonumber \\
    &= f D^{n+1}_{x_{(i,j-\frac{1}{2})}} + (\Delta y - f) \left[\frac{\epsilon_2}{\epsilon_2 n_x^2+\epsilon_1 n_y^2}D^{n+1}_{x_{(A,\Omega_1)}}
       + \frac{(\epsilon_1-\epsilon_2)n_xn_y}{\epsilon_2 n_x^2+\epsilon_1 n_y^2}D^{n+1}_{y_{(A,\Omega_2)}}\right] \nonumber \\
    &= \frac{f(\epsilon_2 n_x^2+\epsilon_1 n_y^2) + (\Delta y- f)\epsilon_2}{\epsilon_2 n_x^2+\epsilon_1 n_y^2} D^{n+1}_{x_{(i,j-\frac{1}{2})}}
       + \frac{(\epsilon_1-\epsilon_2)n_xn_y}{\epsilon_2 n_x^2+\epsilon_1 n_y^2}(\Delta y- f)D^{n+1}_{y_{(C)}}
    \label{newdisAmpere'sdn1}
\end{align}
Similarly,
\begin{align}
    \int_O^C D^{n}_{x(i,y)}dy & = \frac{f(\epsilon_2 n_x^2+\epsilon_1 n_y^2) + (\Delta y- f)\epsilon_2}{\epsilon_2 n_x^2+\epsilon_1 n_y^2}
    D^n_{x_{(i,j-\frac{1}{2})}} \nonumber \\
    & \quad + \frac{(\epsilon_1-\epsilon_2)n_xn_y}{\epsilon_2 n_x^2+\epsilon_1 n_y^2}(\Delta y- f)D^n_{y_{(C)}}
    \label{newdisAmpere'sdn}
\end{align}
Furthermore,
\begin{align}
    D^{n+1}_{y_{(C)}} & = D^n_{y_{(C)}}+\frac{\partial D_y}{\partial t}\rvert_C^{t_n} \cdot \Delta t \nonumber \\
                      & = D^n_{y_{(C)}}-\frac{\partial H_z}{\partial x}\rvert_C^{t_{n+\frac{1}{2}}} \cdot \Delta t \nonumber \\
     &= D^n_{y_{(C)}}-\left(\frac{H^{n+\frac{1}{2}}_{z_{(i,j)}}-H^{n+\frac{1}{2}}_{z_{(i-1,j)}}}{\Delta x}\right) \Delta t
     \label{newdisAmpere'sen1}
\end{align}
By putting \eqref{newdisAmpere'sdn1}, \eqref{newdisAmpere'sdn} and
\eqref{newdisAmpere'sen1} into \eqref{newdisAmpere's}, we obtain
\begin{align}
    \frac{f(\epsilon_2 n_x^2+\epsilon_1 n_y^2) + (\Delta y- f)\epsilon_2}{\epsilon_2 n_x^2+\epsilon_1 n_y^2}
    \left(D^{n+1}_{x_{(i,j-\frac{1}{2})}}-D^{n}_{x_{(i,j-\frac{1}{2})}}\right)
    =\Delta t\left(H^{n+\frac{1}{2}}_{z_{(i,j)}}-H^{n+\frac{1}{2}}_{z_{(i,j-1)}}\right) \nonumber \\
    + \frac{\Delta t}{\Delta x} \frac{(\epsilon_1-\epsilon_2)n_xn_y}{\epsilon_2 n_x^2+\epsilon_1 n_y^2}(\Delta y-f)
    \left(H^{n+\frac{1}{2}}_{z_{(i,j)}}-H^{n+\frac{1}{2}}_{z_{(i-1,j)}}\right)
\end{align}
which, based on $D_x=\epsilon_1 E_x$ in $\Omega_1$, becomes
\begin{align}
    E^{n+1}_{x_{(i,j-\frac{1}{2})}} &= E^{n}_{x_{(i,j-\frac{1}{2})}}
    + \Delta t \frac{\frac{\epsilon_2}{\epsilon_1} n_x^2+ n_y^2}{f(\epsilon_2 n_x^2+\epsilon_1 n_y^2) + (\Delta y- f)\epsilon_2}
                                        \left(H^{n+\frac{1}{2}}_{z_{(i,j)}}-H^{n+\frac{1}{2}}_{z_{(i,j-1)}}\right) \nonumber \\
    &+ \frac{\Delta t}{\Delta x} \frac{(\Delta y-f)(1-\frac{\epsilon_2}{\epsilon_1})n_xn_y}{f(\epsilon_2 n_x^2+\epsilon_1 n_y^2) + (\Delta y- f)\epsilon_2}
    \left(H^{n+\frac{1}{2}}_{z_{(i,j)}}-H^{n+\frac{1}{2}}_{z_{(i-1,j)}}\right)
\label{newdisAmpere'slaw1}
\end{align}
which shows improved discretization of Ampere's law for updating
$E_x.$

It should also be noted that the last term on the right hand side of
\eqref{newdisAmpere'slaw1} have been ignored by CP-EP for algorithm
simplification but it indeed provides necessary corrections for the
interface conditions across multiple media. Furthermore, the
numerical tests have also verified that in order to make the
algorithm for Ampere's Law more accurate, these two $H_z$ values in
the last term should stay close to the interface. That is, if the
normal direction of the interface $n$ satisfies $n_x\cdot n_y
> 0$ as shown in Figure~\ref{amex}, we use
$(H^{n+\frac{1}{2}}_{z_{(i,j)}}-H^{n+\frac{1}{2}}_{z_{(i-1,j)}})$
since these two $H_z$ points are closer to the interface, while for
the other $n_x\cdot n_y\leq 0$ case, we use
$(H^{n+\frac{1}{2}}_{z_{(i+1,j)}}-H^{n+\frac{1}{2}}_{z_{(i,j)}})$ in
\eqref{newdisAmpere'slaw1} instead since these two $H_z$ nodes are
closer to the interface than the other pair. This improvement is
feasible since $H_z$ in continuous across the interface.
\begin{figure}[ht!]
    \begin{center}
            \includegraphics[scale=0.2]{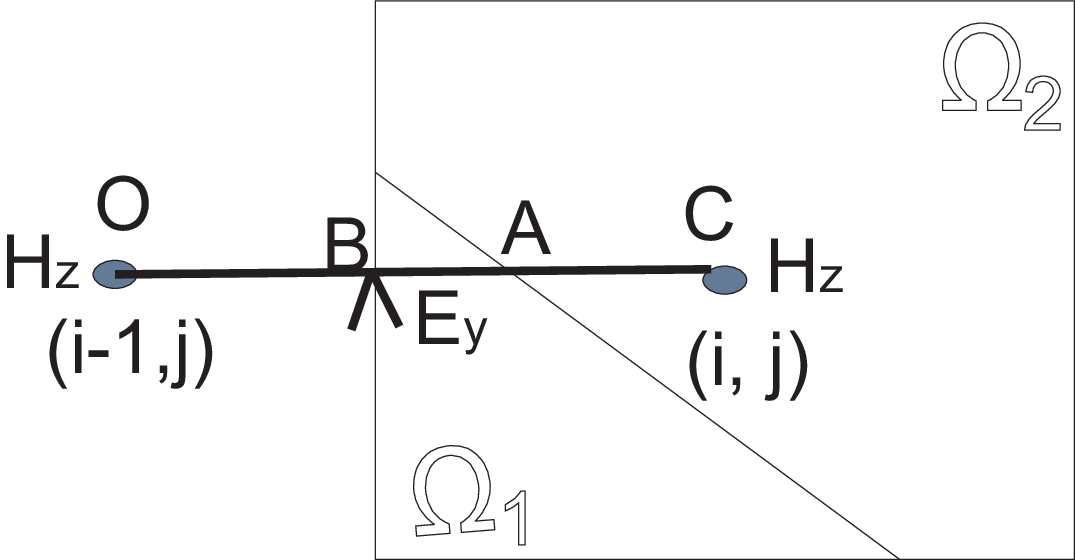}
                \caption{Discretization of the integral along $OC$}
                \label{amey}
    \end{center}
\end{figure}

Similarly, update of $E_y$ at point $B=(i-\frac{1}{2},j)$ by
Ampere's law \eqref{Ampere's} can be taken care analogously as shown
in Figure~\ref{amey}, where the interface $\Gamma$ intersects the
line segment $OC$, projection of $S$ at $xy$-plane, at $A$. Let
$OA=d$. By repeating the same procedure as above while based on
formula \eqref{dintroDy}, we can also obtain the discretization of
Ampere's law for $E_y$ as below:
\begin{align}
    E^{n+1}_{y_{(i,j-\frac{1}{2})}} &= E^{n}_{y_{(i,j-\frac{1}{2})}}
    + \Delta t \frac{\frac{\epsilon_2}{\epsilon_1} n_y^2+ n_x^2}{d(\epsilon_1 n_x^2+\epsilon_2 n_y^2) + (\Delta x- d)\epsilon_2}
                               \left(H^{n+\frac{1}{2}}_{z_{(i-1,j)}}-H^{n+\frac{1}{2}}_{z_{(i,j)}}\right) \nonumber \\
    &+ \frac{\Delta t}{\Delta y} \frac{(\Delta x-d)(1-\frac{\epsilon_2}{\epsilon_1})n_xn_y}{d(\epsilon_1 n_x^2+\epsilon_2 n_y^2) + (\Delta x- d)\epsilon_2}
                               \left(H^{n+\frac{1}{2}}_{z_{(i,j-1)}}-H^{n+\frac{1}{2}}_{z_{(i,j)}}\right)
\label{newdisAmpere'slaw2}
\end{align}
where
$(H^{n+\frac{1}{2}}_{z_{(i,j-1)}}-H^{n+\frac{1}{2}}_{z_{(i,j)}})$ in
the last term will be replaced by
$(H^{n+\frac{1}{2}}_{z_{(i,j)}}-H^{n+\frac{1}{2}}_{z_{(i,j+1)}})$
when\textbf{} $n_x\cdot n_y\leq 0$ holds.

Secondly, let's discretize Faraday's law \eqref{Faraday's} around
the interface $\Gamma$.
\begin{figure}[ht!]
    \begin{center}
            \includegraphics[scale=0.2]{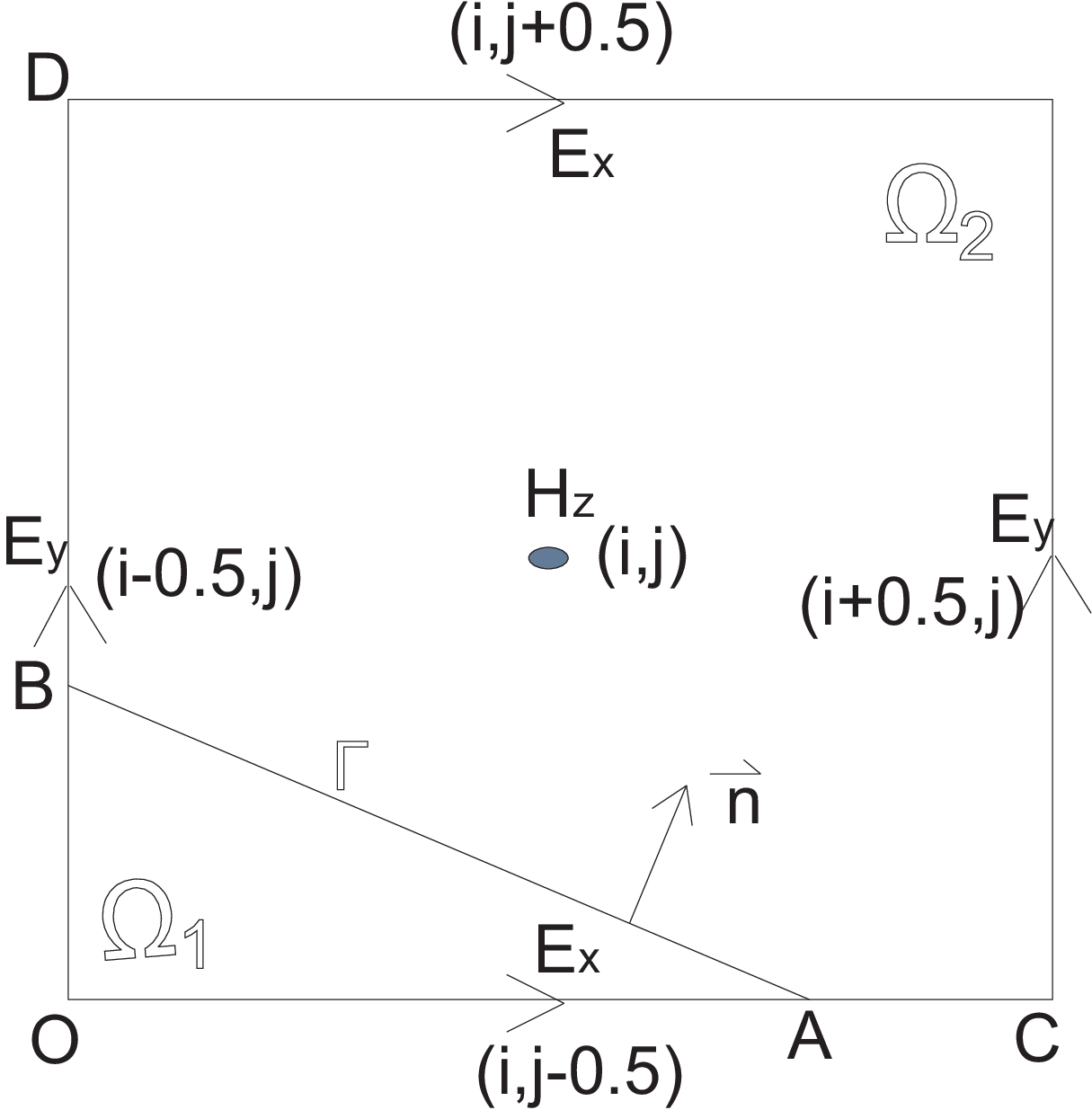}
                \caption{Discretization of Faraday's law across $\Gamma$}
                \label{fahz}
    \end{center}
\end{figure}
We choose $S$ to be the rectangle centered at $(i,j)$. If the medium
interface $\Gamma$ does not cut through interior of $S$, the
standard Yee scheme should apply without modification. So we only
consider the case where $\Gamma$ intersects with $S$. It's obvious
that $\Gamma$ will only intersect with at most two sides, e.g. the
bottom and left side as in Figure~\ref{fahz}. All other cases can be
handled in the same way. Similarly as for Ampere's law, all the
higher-order terms of dx and dy are ignored for the clarification of
the formulas. Discretization of the left hand side term of
\eqref{Faraday's} at the time $t_n$ via finite difference scheme
yields
\begin{align*}
   \frac{ \iint_S \mathbf{H}^{n+\frac{1}{2}} \cdot \mathbf{n} dS - \iint_S \mathbf{H}^{n-\frac{1}{2}} \cdot \mathbf{n} dS  }{\Delta t}
   = -\oint_{\partial S}  \mathbf{E}^n \cdot d\mathbf{l}
\end{align*}
or
\begin{align}
   \iint_S \mathbf{H}^{n+\frac{1}{2}} \cdot \mathbf{n} dS= \iint_S \mathbf{H}^{n-\frac{1}{2}} \cdot \mathbf{n} dS
   -\Delta t \oint_{\partial S}  \mathbf{E}^n \cdot d\mathbf{l}
   \label{newhupdate}
\end{align}
since $\mathbf{H}$ is always perpendicular to the incident
$xy$-plane and so is always continuous across the interface, then by
using midpoint rule,
\begin{align}
   \iint_S \mathbf{H}^{n+\frac{1}{2}} \cdot \mathbf{n} dS= H^{n+\frac{1}{2}}_{z_{(i,j)}} \Delta x \Delta y
   \label{hmid1}
\end{align}
\begin{align}
   \iint_S \mathbf{H}^{n-\frac{1}{2}} \cdot \mathbf{n} dS= H^{n-\frac{1}{2}}_{z_{(i,j)}} \Delta x \Delta y
   \label{hmid2}
\end{align}
However, the situation for $\mathbf{E}$ is much more complicated. In
Figure~\ref{fahz}, the line integral of $\mathbf{E}^n$ along the
right and top sides can be handled as Standard Yee scheme since they
are completely inside $\Omega_2$. But the line integrals along the
bottom and left sides need to be handled separately.
 \begin{align}
   \oint_{\partial S}  \mathbf{E}^n \cdot d\mathbf{l} &= \int_{bottom}  \mathbf{E}^n \cdot d\mathbf{l} +\int_{left}  \mathbf{E}^n \cdot d\mathbf{l}
   \nonumber \\   & \quad+\int^{(j+\frac{1}{2})\Delta y}_{(j-\frac{1}{2})\Delta y}E^n_{(i+\frac{1}{2},y)}dy-\int^{(i+\frac{1}{2})\Delta x}_{(i-\frac{1}{2})\Delta x}E^n_{(x,j+\frac{1}{2})}dx \nonumber \\
&=\int^C_O E^n_{(x,j-\frac{1}{2})}dx-\int^D_O E^n_{(i-\frac{1}{2},y)}dy +E^n_{y_{(i+\frac{1}{2},j)}}\Delta y -E^n_{x_{(i,j+\frac{1}{2})}} \Delta x
\label{cire}
\end{align}
Suppose the portion OA of the bottom side in $\Omega_1$ has length
$d$, and the portion OB of the left side in $\Omega_1$ has length
$f$ as in Figure~\ref{fahz}, then the first two terms on the right
hand side of \eqref{cire} can be handled as below. For the first
term on the right hand side of \eqref{cire}, we have
\begin{align}
    \int^C_O E^n_{(x,j-\frac{1}{2})}dx &= \int^A_O E^n_{(x,j-\frac{1}{2})}dx+\int^C_A E^n_{(x,j-\frac{1}{2})}dx \nonumber \\
    &= E^n_{x_{(i,j-\frac{1}{2})}} \cdot d+ E^n_x \rvert_{A, \Omega_2} \cdot (\Delta x -d)
    \label{rcire}
\end{align}
By using formula \eqref{eintro} with $\Omega_1$ and $\Omega_2$
interchanged, we have
\begin{align}
    E^n_x \rvert_{A, \Omega_2} = (\textstyle{\frac{ \epsilon_1}{ \epsilon_2}}n_x^2+n_y^2)E^n_x \rvert_{A, \Omega_1}+( \frac{ \epsilon_1}{ \epsilon_2} -1)n_xn_yE^n_y \rvert_{A, \Omega_1}
\end{align}
Putting it into \eqref{rcire} yields
\begin{align}
    \int^C_O E^n_{(x,j-\frac{1}{2})}dx &= \textstyle E^n_{x_{(i,j-\frac{1}{2})}} \cdot d+  (\frac{ \epsilon_1}{ \epsilon_2}n_x^2+n_y^2)(\Delta x -d)E^n_x \rvert_{A, \Omega_1} \nonumber \\
    & \textstyle + (\frac{\epsilon_1}{\epsilon_2} -1)n_xn_y(\Delta x -d)E^n_y \rvert_{A, \Omega_1} \nonumber \\
    & \textstyle = [d+(\Delta x-d)(\frac{ \epsilon_1}{\epsilon_2}n_x^2+n_y^2)]E^n_{x_{(i,j-\frac{1}{2})}} \nonumber \\
    & \textstyle + (\frac{\epsilon_1}{\epsilon_2} -1)n_xn_y(\Delta x -d)E^n_{y_{(i,j-\frac{1}{2})}}
    \label{ebottom}
\end{align}
where $E^n_{y_{(i,j-\frac{1}{2})}}$ is the value of $E^n_y$
evaluated at point $(i,j-\frac{1}{2})$ in $\Omega_1$. However, only
$E_x$ instead of $E_y$ is calculated at the mesh point
$(i,j-\frac{1}{2})$. Therefore, we will take the average value of
$E^n_y$ at the neighboring pair of mesh points $(i-\frac{1}{2},j-1)$
and $(i+\frac{1}{2},j)$ if they are both located within $\Omega_1$,
or take the average value of $E^n_y$ at the other neighboring pair
mesh points $(i-\frac{1}{2},j)$ and $(i+\frac{1}{2},j-1)$ if they
are both located within $\Omega_1$. If none of the pairs are located
within $\Omega_1$ , this term will be set to zero as in CP-EP.
Therefore, BC-EP will try to make corrections for the missing terms
in CP-EP if possible.

Similarly, we can calculate the second term on the right hand side
of \eqref{cire}. Since the $E_y$ node $(i-\frac{1}{2},j)$ is located
within $\Omega_2$ instead of $\Omega_1$, so the new $\mathbf{n}$
should be  negative of the old $\mathbf{n}$ used above. Therefore,
applying integration midpoint rule, Taylor expansion and formula
\eqref{eintro} where $(n_x,n_y)$ replaced by $(-n_x,-n_y)$ and $
\epsilon_1, \epsilon_2$ interchanged, we have
\begin{align}
    \int^D_O E^n_{(i-\frac{1}{2},y)}dy &= [f(n_x^2+\textstyle{\frac{ \epsilon_2}{ \epsilon_1}n_y^2)}+(\Delta  y-f)]E^n_{y_{(i-\frac{1}{2},j)}} \nonumber \\
    & \quad+( \frac{ \epsilon_2}{ \epsilon_1} -1)n_xn_yf E^n_{x_{(i-\frac{1}{2},j)}}
    \label{eleft}
\end{align}
when $E^n_{x_{(i-\frac{1}{2},j)}}$ will be taken care similarly as
$E^n_{y_{(i,j-\frac{1}{2})}}$ from the first term. By putting
\eqref{ebottom}, \eqref{eleft} into \eqref{cire}, we get
\begin{align}
       \oint_{\partial S} \mathbf{E}^n \cdot d\mathbf{l} =& \textstyle \ [d+(\Delta x-d)(\frac{ \epsilon_1}{ \epsilon_2}n_x^2+n_y^2)]E^n_{x_{(i,j-\frac{1}{2})}}
       +E^n_{y_{(i+\frac{1}{2},j)}}\Delta y -E^n_{x_{(i,j+\frac{1}{2})}} \Delta x \nonumber \\
       & \textstyle -[f(n_x^2+\frac{\epsilon_2}{\epsilon_1}n_y^2)+(\Delta  y-f)]E^n_{y_{(i-\frac{1}{2},j)}}
       + (\frac{\epsilon_1}{\epsilon_2} -1)n_xn_y(\Delta x -d) E^n_{y_{(i,j-\frac{1}{2})}} \nonumber \\
       & \textstyle - (\frac{\epsilon_2}{\epsilon_1} -1)n_xn_yf E^n_{x_{(i-\frac{1}{2},j)}}
        \label{newcire}
\end{align}
Finally, by putting \eqref{hmid1}, \eqref{hmid2} and \eqref{newcire}
into \eqref{newhupdate} and dividing both sides by $\Delta x \Delta
y$, we obtain the new discretization of Faraday's law:
\begin{align}
    H^{n+\frac{1}{2}}_{z_{(i,j)}} &=  \textstyle H^{n-\frac{1}{2}}_{z_{(i,j)}}  - \frac{\Delta t}{\Delta x \Delta y}\{[d+(\Delta x-d)
       (\frac{\epsilon_1}{\epsilon_2}n_x^2+n_y^2)]E^n_{x_{(i,j-\frac{1}{2})}}+E^n_{y_{(i+\frac{1}{2},j)}}\Delta y  \nonumber \\
       & \textstyle  \quad -E^n_{x_{(i,j+\frac{1}{2})}} \Delta x - [f(n_x^2+\frac{ \epsilon_2}{ \epsilon_1}n_y^2)+(\Delta  y-f)]E^n_{y_{(i-\frac{1}{2},j)}} \nonumber \\
       & \textstyle \quad + (\frac{\epsilon_1}{\epsilon_2} -1)n_xn_y(\Delta x -d)E^n_{y_{(i,j-\frac{1}{2})}}
                          - (\frac{\epsilon_2}{\epsilon_1} -1)n_xn_yf E^n_{x_{(i-\frac{1}{2},j)}} \}
\label{newdisFaraday'slaw}
\end{align}
where the last two terms are expressed by
\begin{align*}
E^n_{y_{(i,j-\frac{1}{2})}} & =
\begin{cases}
    \frac{1}{2}(E^n_{y_{(i-\frac{1}{2},j-1)}} + E^n_{y_{(i+\frac{1}{2},j)}}) & \mbox{ if } (i-\frac{1}{2},j-1)\in \Omega_1, \quad (i+\frac{1}{2},j)\in \Omega_1 \\
    \frac{1}{2}(E^n_{y_{(i-\frac{1}{2},j)}} + E^n_{y_{(i+\frac{1}{2},j-1)}}) & \mbox{ if } (i-\frac{1}{2},j)\in \Omega_1, \quad (i+\frac{1}{2},j-1)\in \Omega_1 \\
    0                                                                        & \mbox{ otherwise }
\end{cases}
\intertext{ and }
E^n_{x_{(i-\frac{1}{2},j)}} & =
\begin{cases}
    \frac{1}{2}(E^n_{x_{(i-1,j-\frac{1}{2})}} + E^n_{x_{(i,j+\frac{1}{2})}}) & \mbox{ if } (i-1,j-\frac{1}{2})\in \Omega_2, \quad (i,j+\frac{1}{2})\in \Omega_2 \\
    \frac{1}{2}(E^n_{x_{(i-1,j+\frac{1}{2})}} + E^n_{x_{(i,j-\frac{1}{2})}}) & \mbox{ if } (i-1,j+\frac{1}{2})\in \Omega_2, \quad (i,j-\frac{1}{2})\in \Omega_2 \\
    0                                                                        & \mbox{ otherwise }
\end{cases}
\end{align*}
If we set $\epsilon_1= \epsilon_2$ in the new algorithm
\eqref{newdisAmpere'slaw1}, \eqref{newdisAmpere'slaw2} for Amepre's
law and \eqref{newdisFaraday'slaw} for Faraday's law, we can
retrieve the standard FDTD scheme in homogeneous medium. Therefore,
BC-EP is a simple extension of the original schemes in order to take
care of the media interface. Meanwhile, the extra terms added in the
formulas will greatly improve the accuracy compared with CP-EP.
Furthermore, BC-EP has a very simple structure and can be merged
into any FDTD software by just revising the effective permittivities
and adding some extra terms when necessary.

It should be mentioned that due to complicated expressions of those
ignored higher-order terms in the newly derived formulas, it is
difficult to demonstrate the convergence order of BC-EP
theoretically. However, subsequent numerical results will show that
BC-EP can achieve the highest convergence compared with other
methods. Another point is that it is still an open challenge to
accurately handle objects with sharp corners, where the resulting
field singularities are known to degrade the accuracy of all
numerical schemes \cite{Oskooi2009}. Therefore, just as other
methods, BC-EP performs well for straight or curved interfaces
instead of interfaces with corners.

\section{Numerical Results}

In this section, we will demonstrate convergence order and stability
of BC-EP numerically. To accomplish this task, we implement the
algorithm to solve the Maxwell's equations on a dielectric cylinder,
together with the staircasing, V-EP, and CP-EP methods. All the
implementations are set up for the 2D FDTD TE case. The total
scattering cross sections (SCS) are calculated by all four methods
and then compared with the well-known analytic solution by Mie
Theory \cite{Bohren1983} so as to measure the accuracy of each
method.
\begin{figure}[ht!]
    \begin{center}
    \includegraphics[scale=0.5]{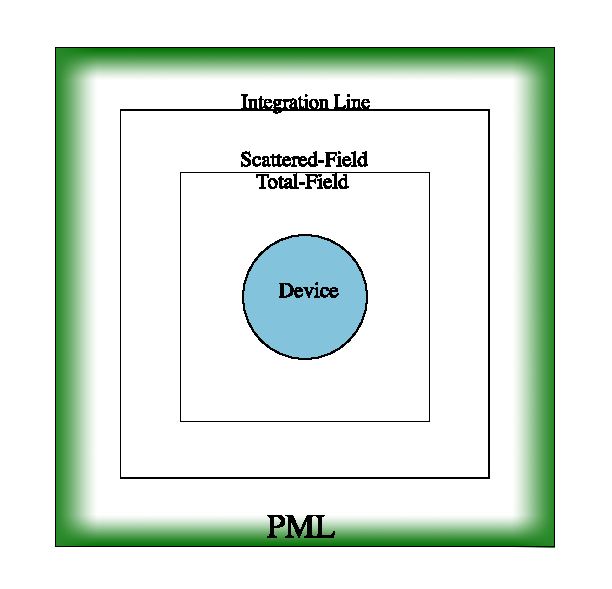}
    \end{center}
    \caption{The numerical test setup using a dielectric cylinder, giving enough space for the total-field/scattered-field and integration line before the PML}
    \label{numericalsetup}
\end{figure}
The dielectric cylinder is simulated in an area that is $10$ times
the radius of the cylinder plus some extra space for PML condition.
The cylinder is centered at $((\frac{N_x}{2}+0.5)\Delta
x,(\frac{N_y}{2}+0.5)\Delta y)$ where $N_x$ and $N_y$ are the number
of the cells along the $x$ and $y$ directions, $\Delta x$ and
$\Delta y$ are the horizontal and vertical mesh sizes of each cell
as seen in Figure~\ref{numericalsetup}. We run the test for two
cases: one case for a larger cylinder with radius $r=400\textrm{nm}$
and another case for a smaller cylinder with $r=150\textrm{nm}$. The
total-field/scattered-field line is three times the length of the
cylinder radius $r$ away from the center, and the integration line
for calculating SCS is four times $r$ away from the center, while
the PML starts at five times $r$ away from the center.

The source wave is a planar wave in the $E_y$ direction with a
Gaussian envelope $e^{\frac{-(t-t_0)^2}{2}\cdot
\frac{1}{2\pi(c/\lambda_0-c/\lambda_1)^2}}\cdot \cos(2\pi ct)$,
where $t_0$ is six times $\Delta t$, $c$ is speed of light in a
vacuum, and $[\lambda_0, \lambda_1]$ is the testing range of
wavelengths. In our simulation, we calculate SCS for $601$
wavelengths ranging from $\lambda_0=400\textrm{nm}$ to
$\lambda_1=1000\textrm{nm}$ of visual light spectrum with equal
distance. Fast Fourier transform (FFT) is then applied to transform
the electric and magnetic fields from spatial domain into light
frequency domain so as to calculate SCS over all related
wavelengths.

We use square unit cells with $\Delta x= \Delta y$ to simplify the
simulation. The time step is set to $\Delta t= \frac{S \Delta x}{c}$
where $S=\frac{0.98}{ \sqrt{3}}$ to ensure stability\
\cite{Taflove2005}. Also for stability concerns, $\Delta x$ were
chosen to divide the smallest tested wave length
($\lambda_0=400\textrm{nm}$) by at least twenty times. The total SCS
is calculated by using the Poynting Vector along with the
integration line as seen in Figure~\ref{numericalsetup}. Meanwhile,
Mie theory is used to calculate the true value of total SCS for each
wavelength. The number of iterations for the main FDTD loop is set
large enough in order to give enough time for the electric and
magnetic waves to leave the simulated area for stable SCS
calculation.
\begin{figure}[htb]
  \centering
  \subfloat[Overall SCS]{\includegraphics[width=0.5\textwidth]{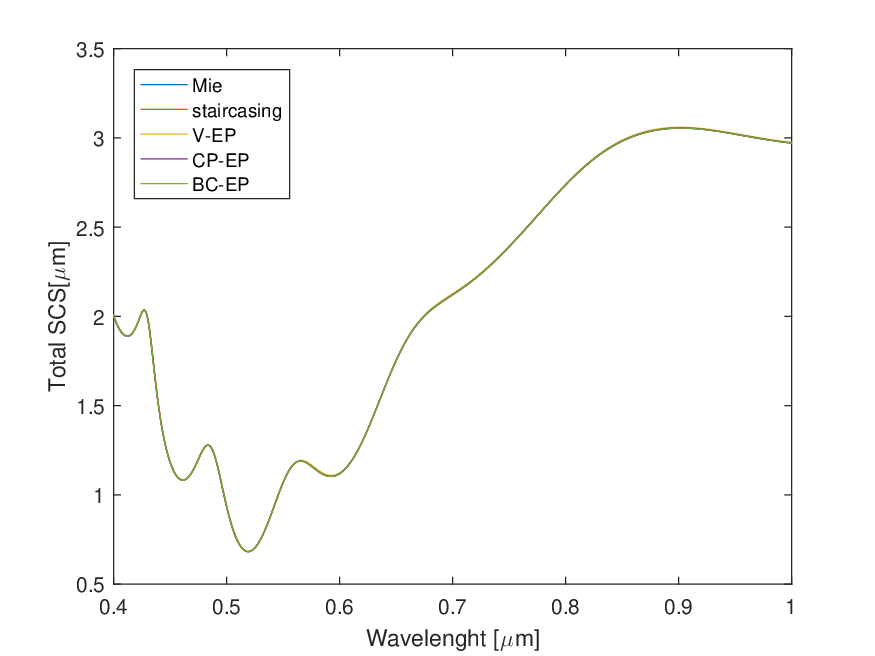}}
  \hfill
  \subfloat[Relative error]{\includegraphics[width=0.5\textwidth]{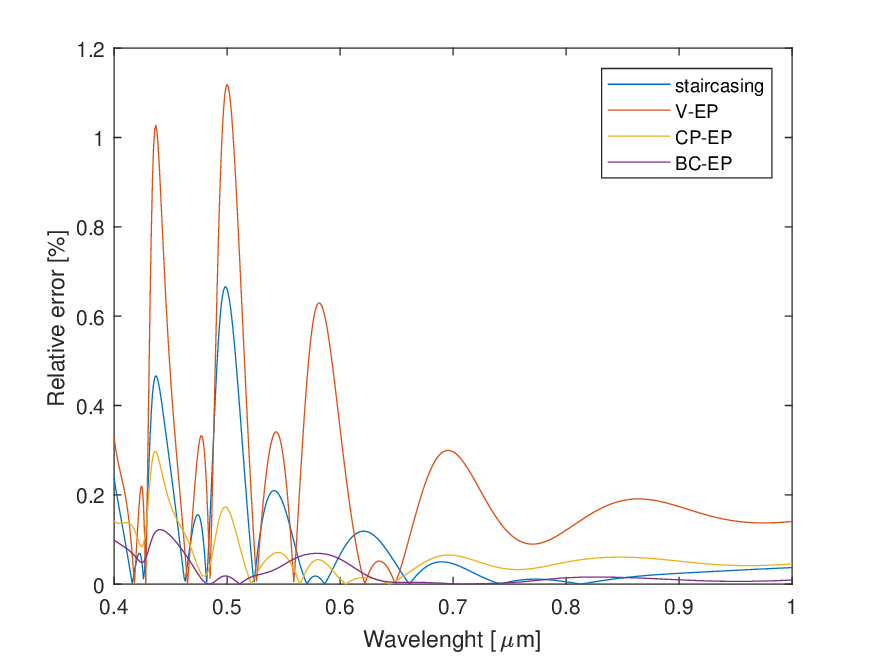}}
  \caption{Total SCS calculation of a cylinder, $\epsilon = 3$ and radius $r = 400\textrm{nm}$. The grid size is $\Delta x=\Delta y = 2.8\textrm{nm}.$}
   \label{simulation3}
\end{figure}

\begin{figure}[htb]
    \begin{center}
            \includegraphics[scale=0.5]{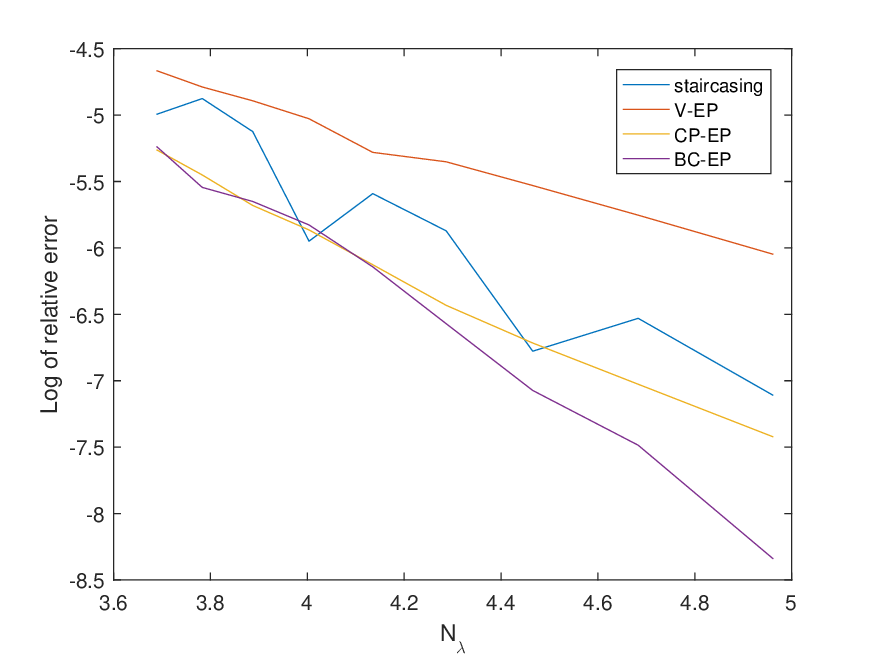}
    \end{center}
    \caption{The average relative error of a cylinder in log scale, $\epsilon = 3$ and radius $r = 400\textrm{nm}$. $N_{\lambda} =\log(\frac{400\textrm{nm}}{\Delta x}).$}
    \label{researchtest3}
\end{figure}

Firstly, we investigate the accuracy and the convergence order of
BC-EP and compare it with other similar methods. We test all the
methods for a circular cylinder with radius $r=400\textrm{nm}$ and
permittivity $\epsilon = 3$. The background media ia always set as
air with $\epsilon=1$. The mesh size is originally set to $\Delta
x=\Delta y = 10\textrm{nm}$. Each method has been run to calculate
the total SCS with different mesh sizes. As seen in (A) of
Figure~\ref{simulation3}, the numerical values from all methods
match with the true solution very well. But we can still see that
BC-EP has the smallest relative error while V-EP has the largest
relative error on most wavelengths shown in (B) of
Figure~\ref{simulation3}. To make a reasonable comparison, for each
mesh size $\Delta x$, the average relative error of SCS is
calculated among all the wavelengths for each method. In order to
observe the error convergence order clearly, the average relative
errors versus the mesh sizes are converted into a log scale where
$N_{\lambda}$ denotes the log of the number of mesh points for the
radius of the cylinder, as seen in Figure~\ref{researchtest3}.

\begin{figure}[htb]
    \begin{center}
            \includegraphics[scale=0.5]{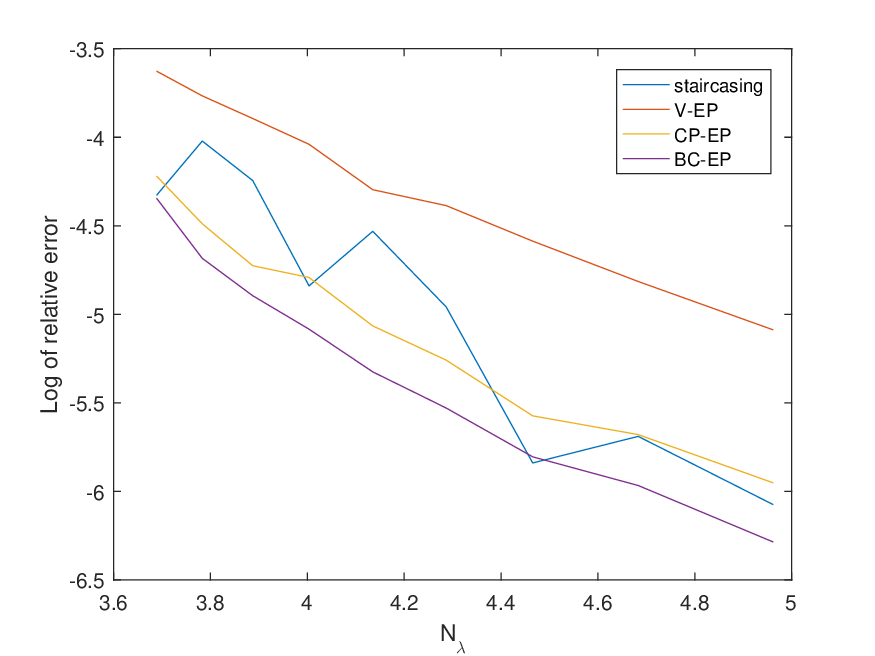}
    \end{center}
    \caption{The average relative error of a cylinder in log scale, $\epsilon = 6$ and radius $r = 400\textrm{nm}$. $N_{\lambda} =\log(\frac{400\textrm{nm}}{ \Delta x}).$}
    \label{researchtest6}
\end{figure}

\begin{figure}[htb]
    \begin{center}
            \includegraphics[scale=0.5]{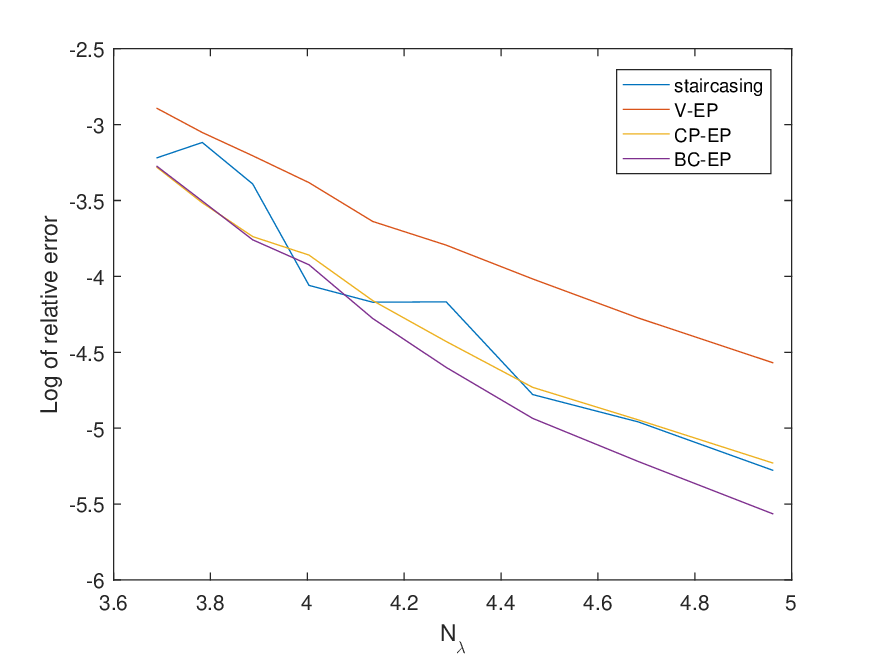}
    \end{center}
    \caption{The average relative error of a cylinder in log scale, $\epsilon = 10$ and radius $r = 400\textrm{nm}$. $N_{\lambda} =\log(\frac{400\textrm{nm}}{ \Delta x}).$}
    \label{researchtest10}
\end{figure}

This whole process is then repeated for other larger permittivity
values $\epsilon=6, 10$. The numerical results for the average
relative errors versus the mesh sizes are reported in
Figure~\ref{researchtest6} for $\epsilon=6$, and in
Figure~\ref{researchtest10} for $\epsilon=10$. Other $\epsilon$
values have similar outcomes.

It can be seen that BC-EP gives the least amount of error for most
cases, followed by CP-EP. Staircasing results are quite erratic and
its error doesn't go down in a uniform manner due to random nature
of some mesh sizes conforming with the cylinder better than others.
V-EP has the worst error but gives a uniform error decrease with
smaller mesh sizes, and thus gives more consistent results over
staircasing.
\begin{table}[htb]
    \centering
    \begin{tabular}{lllll} \hline
        \multicolumn{1}{c}{Permittivity} & \multicolumn{1}{c}{Staircasing} &
        \multicolumn{1}{c}{V-EP} & \multicolumn{1}{c}{CP-EP}  & \multicolumn{1}{c}{BC-EP}\\ \hline \hline
        $\epsilon = 3$ & 1.6621 & 1.0860 & 1.6977 & 2.4386 \\
      %  &&&& \\
       $\epsilon = 4$ & 1.0700 & 1.0850 & 1.4560 & 1.9251 \\
      %  &&&& \\
        $\epsilon = 5$ & 1.2160 & 1.1294 & 1.5267 & 1.6974 \\
      %  &&&& \\
        $\epsilon = 6$ & 1.3723 & 1.1476 & 1.3607 & 1.5247 \\
      %   &&&& \\
         $\epsilon = 7$ & 1.4322 & 1.1834 & 1.3800 & 1.6014 \\
      %  &&&& \\
        $\epsilon = 8$ & 1.4283 & 1.1860 & 1.3564 & 1.6423 \\
      %  &&&& \\
        $\epsilon = 9$ & 1.4867 & 1.2262 & 1.4333 & 1.6837 \\
      %   &&&& \\
        $\epsilon = 10$ & 1.6172 & 1.3183 & 1.5323 & 1.8006 \\ \hline
    \end{tabular}
    \caption{Order of convergence for each FDTD algorithm and for each given permittivity in a cylinder of $r=400\textrm{nm}$.}
    \label{convergence}
\end{table}
To further investigate the order of convergence of all the methods,
we estimate the convergence order by computing the slopes from the
log figures for all the given permittivity. Table~\ref{convergence}
shows BC-EP converges significantly faster than all the other
methods. CP-EP is the second in terms of convergence order and
outperforms both staircasing and V-EP as demonstrated in
\cite{Mohammadi2005} already.

\begin{figure}[htb]
    \begin{center}
            \includegraphics[scale=0.6]{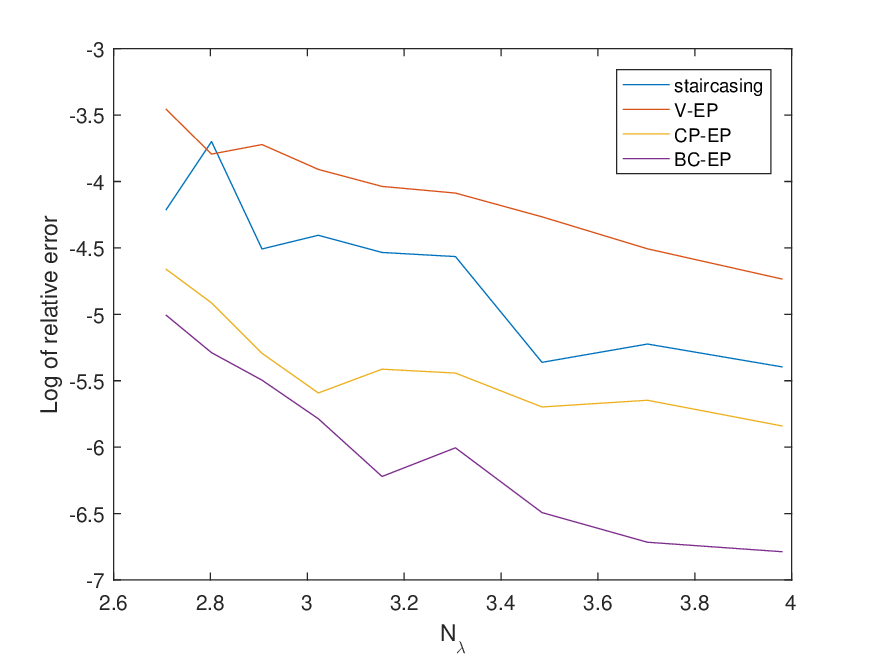}
    \end{center}
    \caption{The average relative error of a cylinder in log scale, $\epsilon = 6$ and radius $r = 150\textrm{nm}$. $N_{\lambda} =\log(\frac{150\textrm{nm}}{\Delta x}).$}
    \label{2researchtest6}
\end{figure}

To complete the final evaluation of the performance for all the
methods, the whole test suite is then repeated for another circular
cylinder with smaller radius $r=150\textrm{nm}$. All the results are
very similar and for brevity, we only report the case of
$\epsilon=6$, as seen in Figure~\ref{2researchtest6}. It can be seen
that BC-EP has a higher convergence order than others. To be more
specific, BC-EP, CP-EP, Staircasing and V-EP have the order of
convergence $1.4005$, $0.92869$, $0.92802$ and $1.0059$,
respectively, based on Figure~\ref{2researchtest6}.

Therefore, BC-EP is more accurate and converges faster than the
other three popular FDTD methods.

\begin{figure}[htb]
  \centering
  \subfloat[$\epsilon = 3$]{\includegraphics[width=0.3\textwidth]{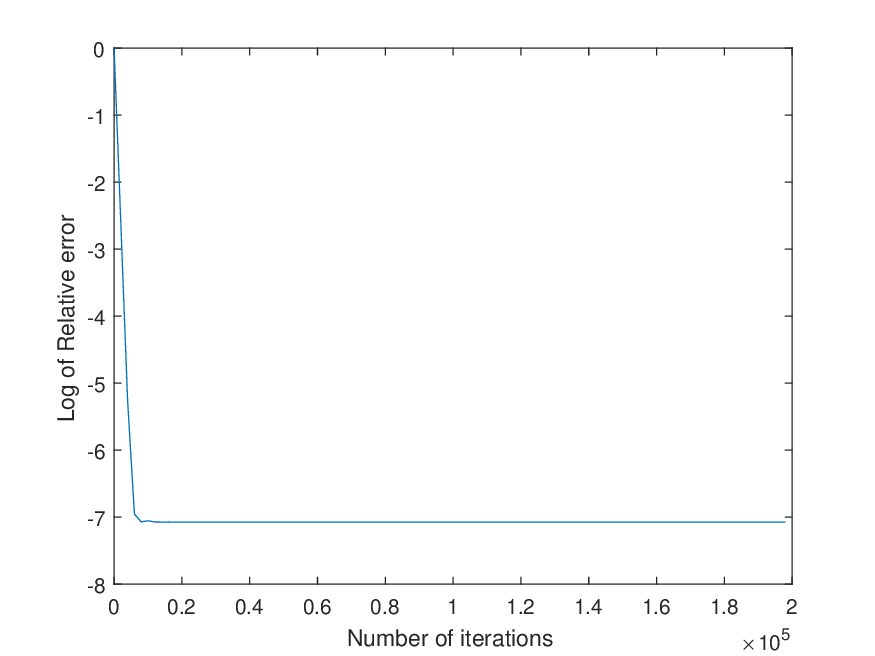}}
  \hfill
  \subfloat[$\epsilon = 10$]{\includegraphics[width=0.3\textwidth]{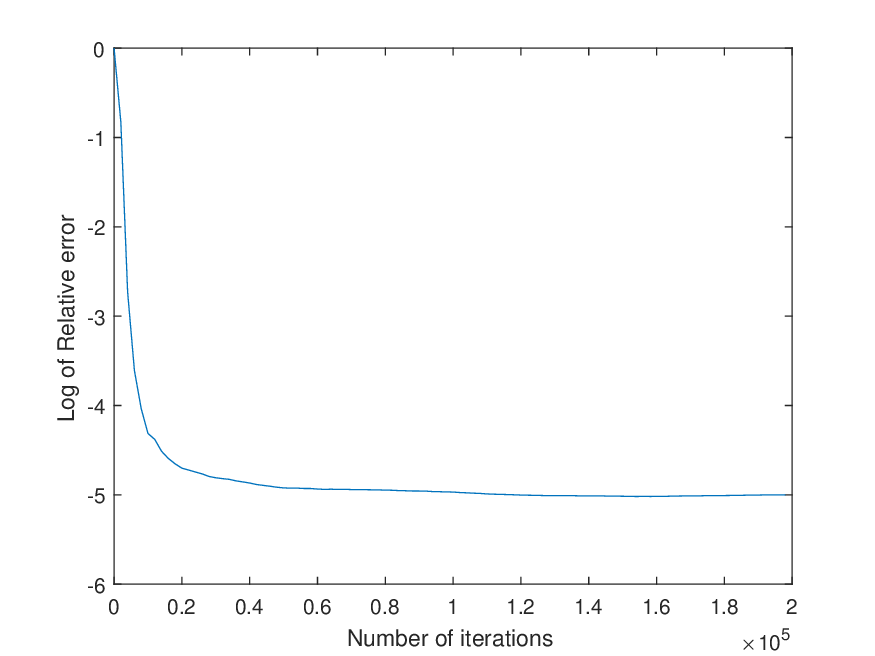}}
  \hfill
  \subfloat[$\epsilon = 30$]{\includegraphics[width=0.3\textwidth]{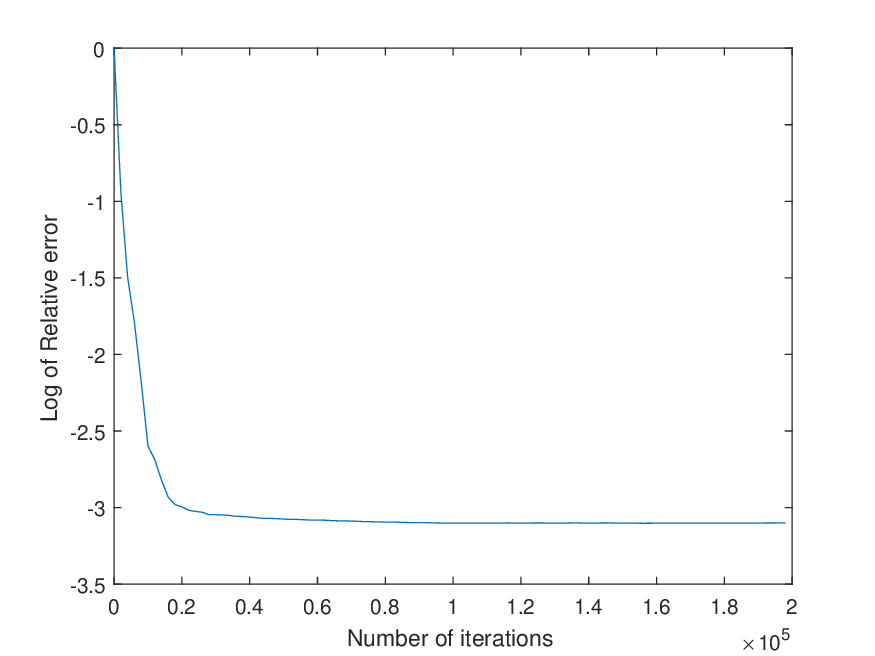}}
  \caption{Relative error in log format for the SCS calculation at a given iteration with $r = 400\textrm{nm}$. $\Delta x=\Delta y=4.6\textrm{nm}.$}
  \label{stable}
\end{figure}

Secondly, we investigate the stability of BC-EP since the stability
of other methods have been demonstrated in the literature. The
algorithm has been run for $200,000$ iterations under FDTD main loop
which is long enough for the electric and magnetic waves to leave
the region completely. This process has been repeated for $\epsilon
= 3, 10, 30$ and for mesh size $\Delta x=10\textrm{nm},
7.3\textrm{nm}, 4.6\textrm{nm},$ and $2.8\textrm{nm}$. At every
$2,000$ iterations, a calculation of the SCS for all $601$
wavelengths is conducted. Then the average relative error between
the calculated SCS at that iteration and the exact SCS over those
wavelengths is calculated. We observe that the result are very
similar for different $\epsilon$ and different $\Delta x$ in the
sense that the relative errors drop down to zero very quickly, as
seen in Figure~\ref{stable}. This shows that there is no
electromagnetic power left in the region and the fields return back
to zero as excepted, meaning no late time instability even at very
high contrast.

\section{Conclusions and Future Research}

We have built a new BC-EP method for the challenging two-dimensional
transverse electric Maxwell's equations involving multiple
dielectric media. Based on a relation for the electric fields across
the dielectric media interface and the integral version of the
Maxwell's equations, we derive this method which represents
Maxwell's equations accurately while the previous numerical methods
skipped some terms from the boundary conditions for the purpose of
algorithm simplification. We then apply BC-EP together with three
other popular FDTD methods, CP-EP, staircasing and V-EP, to solve a
numerical example of a dielectric cylinder with given analytic
solution. The numerical results clearly demonstrate that it always
achieves the highest convergence compared with other methods. The
stability of BC-EP has also been verified numerically. Due to its
simplicity, BC-EP can be merged into other FDTD software packages
easily. The C\texttt{++} source codes for these four methods are
provided as supporting information for public access.

Our next research consideration is to extend BC-EP to solve
two-dimensional Maxwell's equations involving multiple dispersive
media. Numerical performance comparison of this method with other
reported dispersive methods (\cite{Mohammadi2008, Liu2012,
Nguyen2016}) will be provided.

\end{document}